\documentclass[useAMS,usenatbib]{mn2e}
\usepackage{graphicx}

\newcommand{\ms}{$\,$M$_\mathrm{\odot}$}
\newcommand{\ls}{$\,$L$_\mathrm{\odot}$}
\newcommand{\be}{\begin{equation}}
\newcommand{\ee}{\end{equation}}
\newcommand{\stars}{{\sc stars}}
\newcommand{\el}[2]{\ensuremath{^{#1}\mathrm{#2}}}

\title[The progenitor of SN1993J]{Modelling the binary progenitor of supernova 1993J}
\author[R.~J. Stancliffe \& J.~J. Eldridge]{Richard J. Stancliffe$^{1,2}$\thanks{E-mail:
Richard.Stancliffe@sci.monash.edu.au} \& John J. Eldridge$^1$\\
$^1$Institute of Astronomy, The Observatories, Madingley Road, Cambridge CB3 0HA, U.K. \\
$^2$Centre for Stellar and Planetary Astrophysics, School of Mathematics, Building 28, Monash University, Victoria 3800, Australia }
\begin{document}
\bibliographystyle{mn2e}

\date{Accepted 0000 December 00. Received 0000 December 00; in original form 0000 October 00}

\pagerange{\pageref{firstpage}--\pageref{lastpage}} \pubyear{0000}

\maketitle

\label{firstpage}

\begin{abstract}
We have developed a detailed stellar evolution code capable of following the simultaneous evolution of both stars in a binary system, together with their orbital properties. To demonstrate the capabilities of the code we investigate potential progenitors for the Type IIb supernova 1993J, which is believed to have been an interacting binary system prior to its primary exploding. We use our detailed binary stellar evolution code to model this system to determine the possible range of primary and secondary masses that could have produced the observed characteristics of this system, with particular reference to the secondary. Using the luminosities and temperatures for both stars (as determined by \citealt{2004Natur.427..129M}) and the remaining mass of the hydrogen envelope of the primary at the time of explosion, we find that if mass transfer is 100 per cent efficient the observations can be reproduced by a system consisting of a 15\ms\ primary and a 14\ms\ secondary in an orbit with an initial period of 2100 days. With a mass transfer efficiency of 50 per cent, a more massive system consisting of a 17\ms\ primary and a 16\ms\ secondary in an initial orbit of 2360 days is needed. We also investigate some of the uncertainties in the evolution, including the effects of tidal interaction, convective overshooting and thermohaline mixing.
\end{abstract}


\section{Introduction}

The supernova 1993J, which occurred in M81, was an unusual event. Its spectrum initially showed hydrogen lines like a normal Type II supernova, but these subsequently disappeared giving rise to a Type Ib-like spectrum, which shows broad helium lines and no silicon lines \citep[][and references therein]{2000AJ....120.1487M}. The supernova is thus classified as a Type IIb. It was suggested that in order for this transition from a Type II to a Type Ib-like lightcurve to happen, only a small amount of the star's hydrogen envelope could remain at the time of explosion \citep{1993Natur.364..509P,1994ApJ...429..300W}. However, this is difficult to explain in terms of the evolution of single stars -- a star with the observed luminosity of the progenitor of SN1993J would be expected to retain a significant amount of its hydrogen-rich envelope \citep[see e.g.][]{2008ApJ...676.1016D}. It was therefore suggested that the presence of a binary companion was necessary to explain the observations \citep{1993Natur.364..509P}. In this scenario, the red giant progenitor would fill its Roche lobe and be stripped of almost all its envelope.

\citet{1994AJ....107..662A} noted that photometry of the progenitor could not be explained by single star models. Excesses in the ultra-violet and in the B-band could best be fitted with a spectrum which was the composite of a red star together with either a blue star or the net spectrum of an OB association. \citet{2002PASP..114.1322V} suggested that there were nearby blue stars which could account for the blue excess in observations of the progenitor. Subsequently, \citet{2004Natur.427..129M} were able to detect the signature of a massive star at the location of the supernova, proving that the supernova had come from a binary system. SN1993J is therefore the first supernova to have been confirmed to come from a binary system. \citet{2004Natur.427..129M} give the parameters of the primary as $\log L/\mathrm{L_\odot} = 5.1\pm0.3$ and $\log T_\mathrm{eff}/\mathrm{K} = 3.63\pm0.05$, while they identify the secondary as having the parameters $\log L/\mathrm{L_\odot} = 5.0\pm0.3$ and $\log T_\mathrm{eff}/\mathrm{K} = 4.3\pm0.1$.

Initial attempts to model the progenitor system of SN1993J, along with the supernova light curves that these models would produce, were made by \citet{1994ApJ...429..300W}. These authors concluded that the primary had to be in the mass range 13-16\ms\ and that at the time of explosion the helium core mass was  $4.0\pm0.5$\ms, while the hydrogen-rich envelope was around $0.20\pm0.05$\ms\ based on the bolometric light curve obtained from modelling the explosion of their evolution models.

However, at the time of these models nothing was known about the properties of the secondary. Now that we have measurements of the luminosity and temperature of the secondary, it is fruitful to revisit the modelling of this system to see what new constraints can be added. Particularly, the binary models of \citet{1994ApJ...429..300W} looked at systems where the initial mass ratio of the components (secondary-to-primary mass) was between around 0.6 and 0.75. It is clear from the location of the secondary that the mass ratio was much closer to unity. The model presented by \citet{2004Natur.427..129M} was an improvement on earlier work as these authors had obtained details of the secondary. However, the degeneracy of the solution was not studied. Here we build upon their model by examining how sensitive the system is to the initial conditions and the treatment of  detailed physics such as tides, thermohaline mixing and convective overshooting. 

\section{The stellar evolution code}
We use an extensively modified version of the stellar evolution code \stars, which was originally developed by \citet{1971MNRAS.151..351E} and has been updated by many authors. The last major update was by \citet{1995MNRAS.274..964P} and the version used in this work makes substantial updates to this. Variants of Eggleton's code are unique in that they compute the stellar structure and chemical evolution simultaneously, iterating on all variables {\it at the same time} in order to converge a new model. Most codes converge a structure and then use this to converge the chemical evolution. While this has been shown to be unimportant in most phases of stellar evolution, there are some situations where simultaneous solution may be necessary\footnote{Specifically, if mixing, burning and structural changes proceed on similar timescales. This is independent of whether one uses a Lagrangian mesh or not.} \citep[see the discussion in][for further details]{2006MNRAS.370.1817S}. The code now follows the chemical evolution of the 7 energetically important species \el{1}{H}, \el{3}{He}, \el{4}{He}, \el{12}{C}, \el{14}{N}, \el{16}{O} and  \el{20}{Ne} alongside the structure variables. The inclusion of \el{3}{He} is necessary for the modelling of low mass stars following the discovery of \citet*{2006Sci...314.1580E,2008ApJ...677..581E}, who showed that a mean molecular weight inversion caused by the burning of \el{3}{He}  leads to mixing on the red giant branch. 

Convection is treated within the code using the mixing length theory \citep{1958ZA.....46..108B} and the models were evolved with a mixing length parameter, $\alpha=2.0$. This value is obtained by calibrating to a solar model. Convective overshooting was employed via the prescription of \citet*{1997MNRAS.285..696S} and an overshooting parameter of $\delta_\mathrm{ov}=0.12$ was used\footnote{It should be noted that the overshooting parameter was only calibrated in the mass range 2-7\ms\ \citep{1997MNRAS.285..696S}. We are making the assumption that this value also applies outside this mass range.}. The nuclear reaction rates used in the code were recently updated by \citet{2005MNRAS.360..375S} and \citet{stancliffe05}, as part of an introduction of routines for following the nucleosynthesis of around 40 isotopes from D to \el{32}{S} and important iron group elements.

\subsection{Binary evolution}
The most substantial modification to the code is the incorporation of the ideas from Eggleton's binary evolution code {\sc twin}, \citep{2002ApJ...575..461E}. The {\sc twin} code simultaneously solves the equations for the evolution of both stars in a binary system along with their orbital properties. We have modified the \stars\ code to be able to do this. This necessarily requires the inclusion of the physics of binary interaction. We assume that the orbit of the binary is circular. The extension to elliptical orbits is reserved for future work. For the systems we are interested in the orbit will circularise before the primary fills its Roche lobe.

In order to model the orbits of the stars, we have added three new variables to the code: the angular momentum of the orbit and the spin angular momentum of both stars. We assume that the stars rotate as solid bodies. Recent simulations of stars with both rotation and magnetic fields suggest that this may be a reasonable assumption \citep[at least for main sequence evolution, see][]{2005A&A...440.1041M}. We account for tidal interaction of the binary using the equations of \citet{1981A&A....99..126H} and we follow the prescriptions of \citet*{2002MNRAS.329..897H} for the details of the equilibrium tide with convective damping and the dynamical tide with radiative damping.

The code also accounts for the loss of angular momentum via winds and for the transfer of angular momentum from one star to the other, both with direct mass transfer from Roche lobe overflow and for wind accretion. We account for the variation in composition of the material that is transferred from one star to the other. Again, this is done both for the case of Roche lobe overflow and wind accretion.

We find it necessary to modify the physics of accretion, mass loss and angular momentum loss in the case that a star rotates close to break-up. We do this by modifying the mass-loss:
\be
\dot{M}(\Omega) = {\dot{M}(0)\over 1-\frac{\Omega}{\Omega_\mathrm{crit}}}
\ee
where $\dot{M}(\Omega)$ is the mass-loss rate as a function of rotation rate, $\dot{M}(0)$ is the mass-loss rate in the absence of rotation, $\Omega$ is the angular frequency of rotation and $\Omega_\mathrm{crit}$ is the angular frequency for critical rotation. This increase in $\dot{M}$ reflects the fact that as the star spins more rapidly, its surface layers are less gravitationally bound. Also, we presume that a star which is rotating at close to its critical rotation rate will not continue to accrete more material. We therefore reduce the accretion rate by a factor of $1-\frac{\Omega}{\Omega_\mathrm{crit}}$. Finally, we also enhance the rate of loss of spin angular momentum by a factor of $(1 - \frac{\Omega}{\Omega_\mathrm{crit}})^{-1}$.

During mass transfer, it is not guaranteed that the accreting star will be able to accept all the matter than the primary transfers. We therefore set a maximum rate of accretion for a star equal to the star's mass divided by its Kelvin-Helmholtz timescale, as is done by \citet{2002MNRAS.329..897H}. This is in addition to the factor discussed above.

The code also includes the physics for thermohaline mixing, using the prescription of \citet{1980A&A....91..175K} as implemented by \citet{2007A&A...464L..57S}. Thermohaline mixing is an important process for systems undergoing mass transfer. It occurs when the mean molecular weight of the stellar gas increases towards the surface. Although this situation is dynamically stable as long as the density decreases
outward, a secular instability sets in when a gas element is displaced
downwards and compressed. Such an element must be hotter than the
surrounding gas by virtue of its higher molecular weight and if it
loses heat to its surroundings its density increases and it will
continue to sink. As a result mixing occurs on a thermal timescale,
until the molecular weight difference has disappeared
\citep{1980A&A....91..175K}. In interacting binary systems, mass transfer (by wind accretion or Roche lobe overflow) may lead to the transfer of material that has been processed by nuclear burning from the primary to the surface of the less evolved secondary. This alters the surface composition of the secondary. The extent to which the surface composition is changed depends on how much the transferred matter is mixed with the pristine stellar material.

\subsection{Opacities}
The opacity routines of the code were substantially overhauled by \citet{2004MNRAS.348..201E}. These authors modified the routines so that they now account for changes in the opacity related to variations in the carbon and oxygen composition of the material. The high temperature opacities are taken from \citet{1996ApJ...464..943I}, while at temperatures below $\log_{10} T/\mathrm{K}=4.0$, the opacities of \citet{1994ApJ...437..879A} and \citet{2005ApJ...623..585F} are used. Opacity tables from $Z=0$ up to $Z=0.05$ have been produced. The low temperature opacities used by \citet{2004MNRAS.348..201E} do not account for molecular opacities. These are only important at low temperatures and are particularly important for carbon-rich matter. These conditions are not encountered in the mass ranges studied in this work.\footnote{However, \citet{2008MNRAS.389.1828S} have modified the opacity routines to include the prescription of \citet{2002A&A...387..507M} in order to study the evolution of asymptotic giant branch (AGB) stars.}

\subsection{Mass-loss rates}
The new version of the code now has some of the most commonly used mass loss rates included in it. In addition to the Reimers' rate \citep{1975psae.book..229R}, the code also includes the prescriptions of \citet{1993ApJ...413..641V} and \citet{1995A&A...297..727B}. These rates have recently been employed by \citet{2007MNRAS.375.1280S} and \citet*{2008MNRAS.385..301L} in modelling the evolution of AGB stars.

For the massive stars considered in this work, we employ the mass-loss rates described in \citet{2004MNRAS.353...87E} and \citet{2006A&A...452..295E}. These authors use the rates of \citet*{2001A&A...369..574V} for OB stars, which have a metallicity scaling factor of $(Z/Z_\odot)^{0.69}$, the rates of \citet{1988A&AS...72..259D} for pre-Wolf-Rayet phases of evolution and the rates of \citet{2000A&A...360..227N} for Wolf-Rayet phases. A common metallicity scaling factor of $(Z/Z_\odot)^{0.5}$ is used for both these phases.

\section{Modelling and results}

Based on the luminosity limits of the primary and secondary, we can immediately constrain what mass of star the primary was and what the secondary mass would have been at the point the primary exploded. If the primary is less than 11\ms\ then the star does not become luminous enough to match the lower limit. If it is more massive than about 32\ms\ then the star is too bright to pass through the upper luminosity limit. Similarly, the secondary must lie between about 14\ms\ and 28\ms. A further constraint is imposed on the possible mass ratio of the system. The position of the secondary in the HR diagram tells us that it must have reached (or be extremely close to) the end of its main-sequence lifetime and therefore that the mass ratio of the system must be close to unity.

We have constructed a grid of models of varying initial stellar masses and initial periods. We use mass pairs of 12+11\ms, 15+14\ms, 17+16\ms\ and 20+19\ms\ and initial periods of 1500, 1750, 2000, 2250 and 2500 days. A metallicity of $Z=0.04$ is used as \citet{2003MNRAS.343..735S} give the metallicity as being roughly twice the solar value. The initial abundances used in the models are solar scaled, based on the values from \citet{1989GeCoA..53..197A}. These systems are then evolved up to the end of carbon burning. {\it The star would explode shortly after this point}. We initially assume that the mass transfer during Roche lobe overflow is 50\% efficient. All of the angular momentum lost by the mass-losing star is lost along with this material, so that no angular momentum is transferred to the accreting star during the mass exchange\footnote{The fraction of the angular momentum that is accreted by the mass gaining star presumably depends on the manner in which the material is accreted (e.g. through an accretion disc). As we do not know what this was in our system we treat it as a free parameter. We have run simulations where we allow some angular momentum to be accreted by the mass gaining star and they do not differ substantially from the simulations presented herein.}.

The progenitor of the supernova was identified as a red supergiant. This means that an interaction between the stars had to have occurred when the primary had a convective envelope. This presents us with a problem when dealing with Roche Lobe overflow. Stripping mass from a star with a convective envelope leads to the expansion of the star and to a positive feedback for mass loss. We find that the primary rapidly loses mass and quickly reaches mass-loss rates of $10^{-2}$\ms\,yr$^{-1}$ (this value is in good agreement with the mass-loss rates reported by \citealt{2004Natur.427..129M}). Convergence of the code fails if the mass-loss rate gets much higher. We therefore cap the mass-loss rate at this value, purely for numerical convenience. It is possible that the mass-loss rate in an actual stellar system under these conditions could be higher. 

\begin{figure*}
\includegraphics[width=\columnwidth]{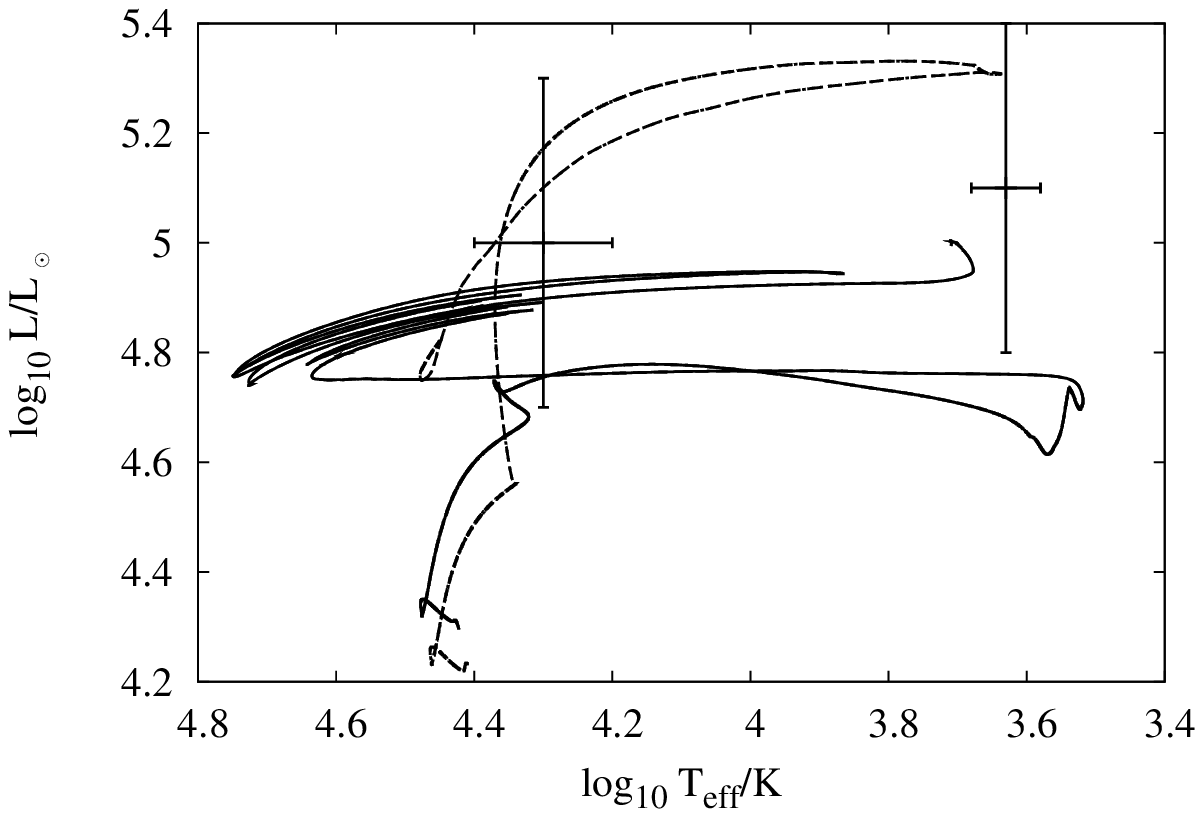}
\includegraphics[width=\columnwidth]{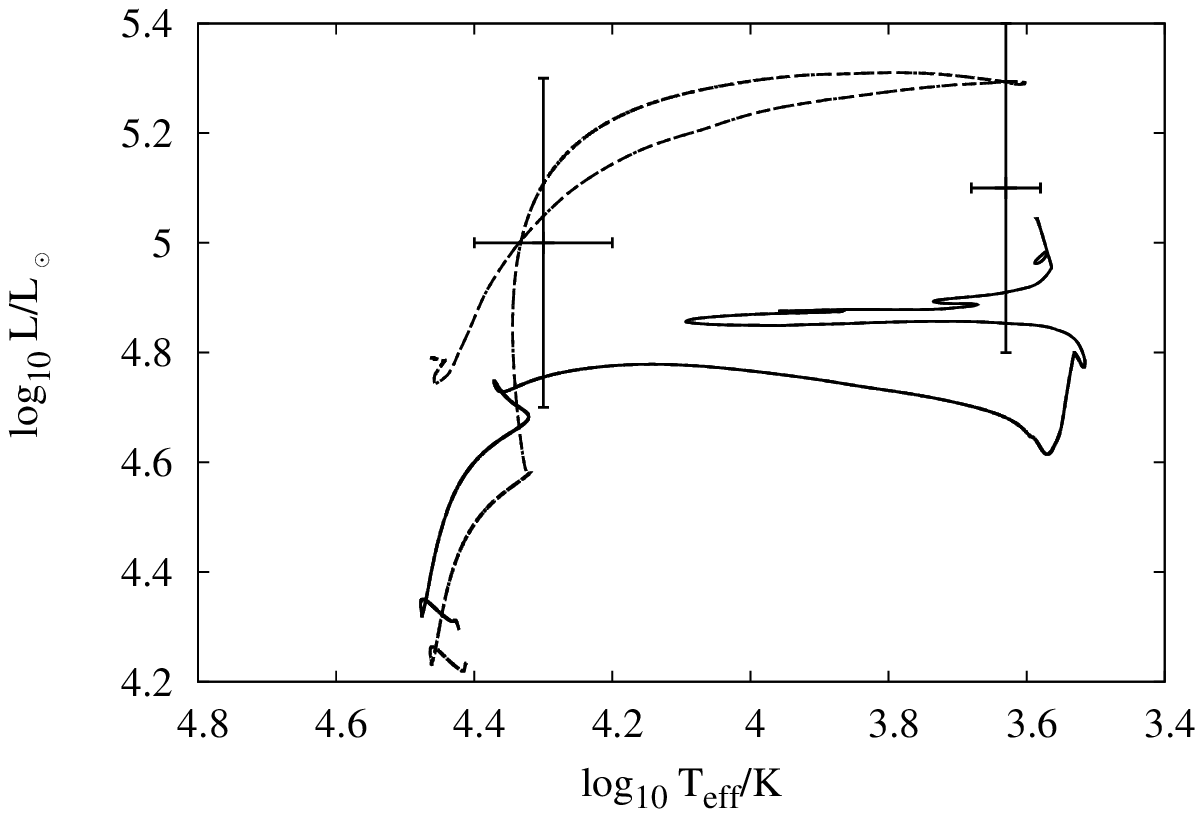}
\includegraphics[width=\columnwidth]{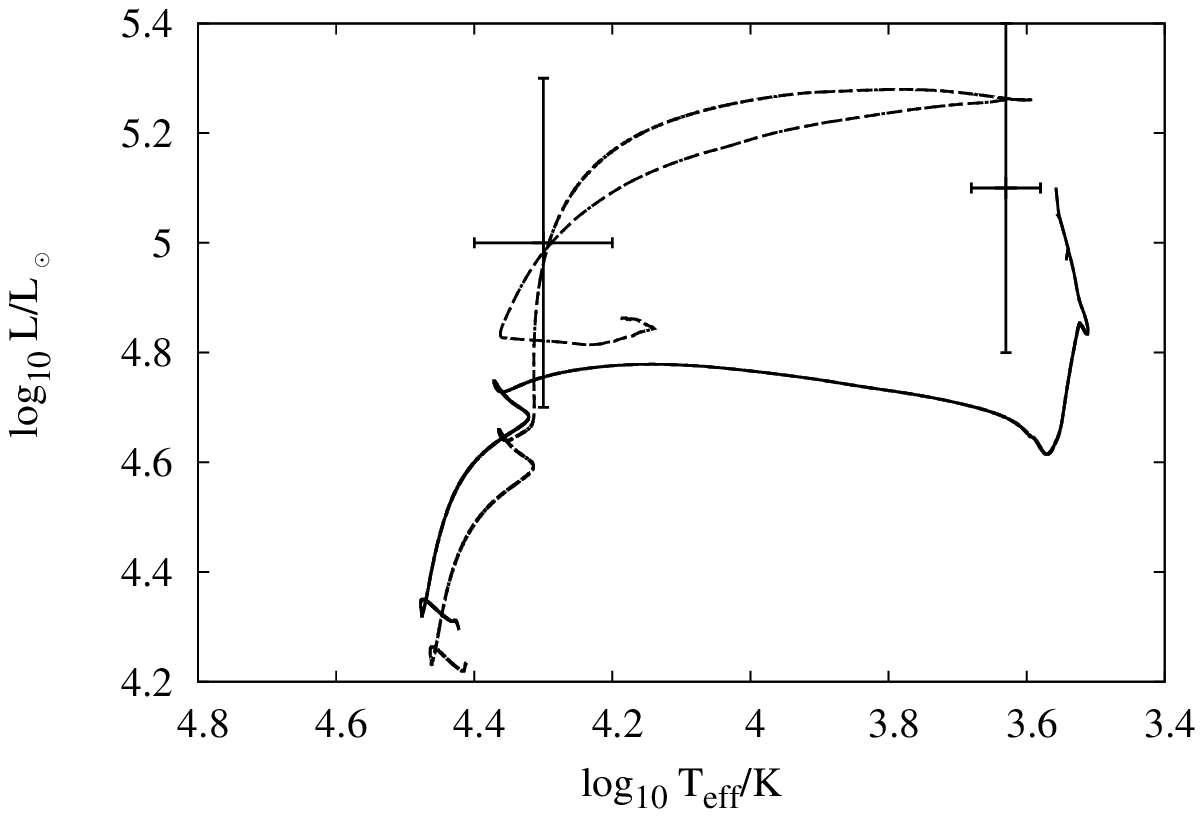}
\includegraphics[width=\columnwidth]{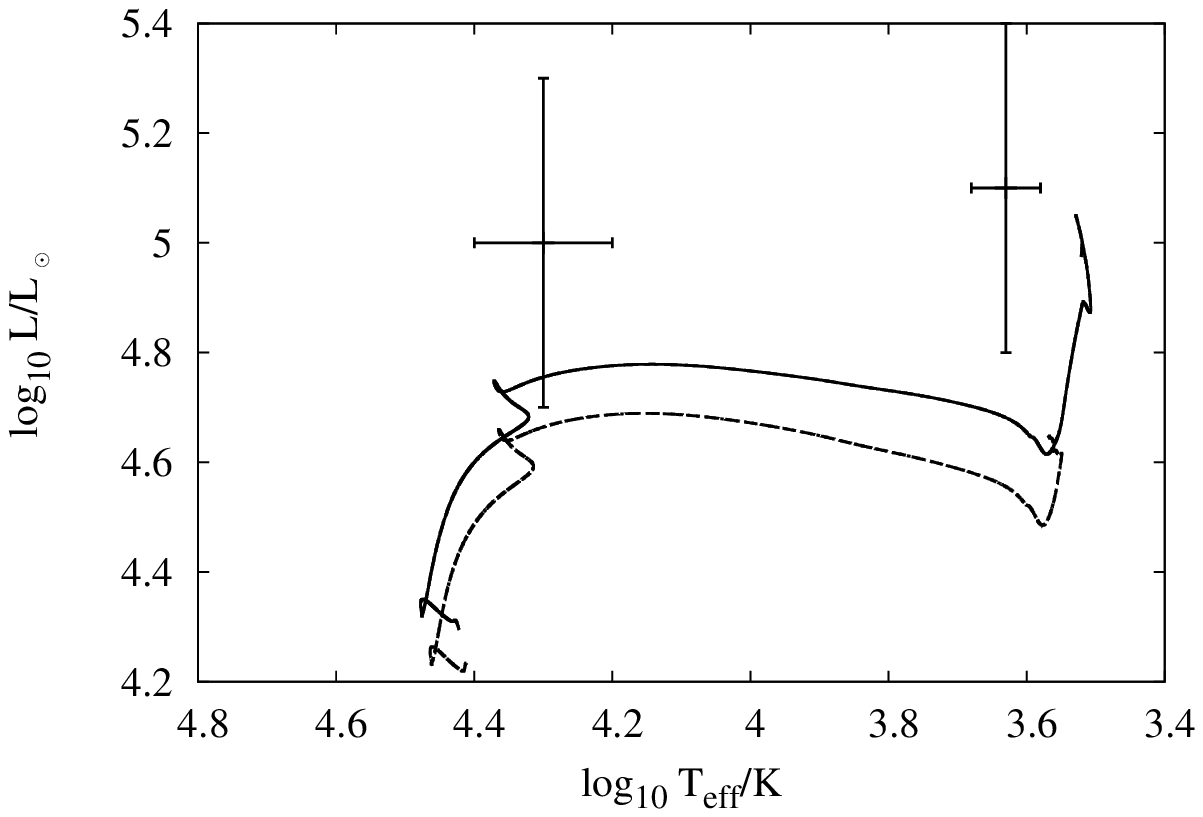}
\caption{Example HR diagrams for systems with a primary of 15\ms\ and a secondary of 14\ms. The initial periods of the systems are: top left, 1500 days; top right, 1750 days; bottom left, 2000 days; bottom right, 2250 days. Note that the model with the 1500 day initial period suffers from breathing pulses at the end of core He burning. The crosses with errorbars denote the observed locations of the primary (at $\log T_\mathrm{eff}/\mathrm{K}=3.63$) and secondary of SN1993J.}
\label{fig:p15s14}
\end{figure*}

Figure~\ref{fig:p15s14} shows the evolution of the 15+14\ms\ systems in the Hertzsprung-Russell (HR) diagram. We find that if the initial period of the system is 2250 days or greater the secondary is too evolved by the time the primary explodes. The model with an initial period of 1500 days encounters some numerical problems when the primary star reaches the end of core helium burning. The oscillations in the HR diagram are breathing pulses \citep[see e.g.][for a discussion of this phenomenon]{1985ApJ...296..204C}. As core helium burning shuts down, the convective core expands slightly, mixing in fresh helium. This prolongs the helium burning. It appears to be a numerical artefact as increasing the resolution on the boundary of the convective core and reducing the timestep size at the end of core helium burning reduces the magnitude of the effect and in some cases removes it altogether.

\begin{table*}
\begin{center}
\begin{tabular}{ccccccccccccc}
\hline
Initial & Initial & M$_\mathrm{final}$ & M$_\mathrm{He}$ & Envelope  & $\log$\ L$^1$/\ls & $\log$\ T$^1_\mathrm{eff}/\mathrm{K}$ & $\log$\ L$^2$/\ls & $\log$\ T$^2_\mathrm{eff}/\mathrm{K}$ & N/C \\
masses & period & & & mass  \\
(\ms) & (days) & (\ms) & (\ms) & (\ms) & & & & & \\ 
\hline
12+11 & 2000 & 4.05 & 3.38 & 0.66 & 4.83 & 3.49 & 4.48 & 4.40 & 13.1 \\ 
12+11 & 2250 & 4.41 & 3.40 & 1.74 & 4.82 & 3.48 & 4.52 & 4.40 & 13.2 \\ 
12+11 & 2500 & 4.93 & 3.40 & 1.91 & 4.84 & 3.47 & 4.61 & 4.37 & 11.0 \\
\hline
15+14 & 1500 & 4.72 & 4.64 & 0.07 & 4.99 & 3.71 & 4.82 & 4.45 & 2.81 \\ 
15+14 & 1750 & 4.94 & 4.85 & 0.09 & 4.93 & 3.57 & 4.79 & 4.46 & 12.2 \\ 
15+14 & 2000 & 5.09 & 4.89 & 0.19 & 5.08 & 3.55 & 4.86 & 4.20 & 7.2 \\ 
15+14 & 2250 & 5.30 & 4.92 & 0.38 & 5.02 & 3.53 & 4.65 & 3.59 & 7.9 \\ 
15+14 & 2500 & 5.59 & 4.94 & 0.65 & 5.04 & 3.51 & 4.64 & 3.56 & 9.7 \\ 
\hline
17+16 & 1500 & 5.15 & 0.00 & 0.04 & 5.08 & 3.69 & 5.00 & 4.50 & 3.8 \\ 
17+16 & 1750 & 5.28 & 0.00 &  0.01 & 5.09 & 3.79 & 4.98 & 4.47 & 2.0 \\ 
17+16 & 2000 & 5.67 & 0.00 & 0.05 & 5.12 & 3.69 & 4.96 & 4.46 & 2.5 \\ 
17+16 & 2250 & 5.94 & 5.73 & 0.20 & 5.14 & 3.56 & 4.94 & 4.48 & 11.4 \\ 
17+16 & 2500 & 6.22 & 5.77 & 0.44 & 5.15 & 3.53 & 4.84 & 3.56 & 5.2 \\ 
\hline
20+19 & 1500 & 5.73 & 0.00 & 0.00 & 5.09 & 4.95 & 5.15 & 4.50 & 7.0 \\
20+19 & 1750 & 6.04 & 0.00 & 0.00 & 4.99 & 5.11 & 5.15 & 4.50 & 9.1 \\ 
20+19 & 2000 & 6.28 & 0.00 & 0.00 & 5.22 & 4.65 & 5.15 & 4.50 & 2.7 \\ 
20+19 & 2250 & 6.60 & 6.58 & 0.02 & 5.24 & 4.22 & 5.15 & 4.50 & 3.3 \\ 
20+19 & 2500 & 6.85 & 6.82 & 0.02 & 5.23 & 3.84 & 5.13 & 4.48 & 3.8 \\ 
\hline
No Tides \\
\hline
12+11 & 1500 & 4.00 & 3.39 & 0.61 & 4.83 & 3.50 & 4.49 & 4.43 & 10.4 \\ 
12+11 & 1750 & 5.12 & 3.42 & 1.70 & 4.78 & 3.48 & 4.55 & 4.39 & 11.4 \\ 
12+11 & 2000 & 5.14 & 3.41 & 1.73 & 4.84 & 3.48 & 4.51 & 4.38 & 10.5 \\ 
12+11 & 2250 & 10.23 & 3.38 & 6.85 & 4.83 & 3.51 & 4.26 & 4.28 & 5.2 \\ 
12+11 & 2500 & 10.24 & 3.41 & 6.83 & 4.78 & 3.51 & 4.26 & 4.28 & 5.2 \\ 
\hline
15+14 & 1500 & 5.09 & 4.88 & 0.21 & 4.85 & 3.56 & 4.81 & 4.48 & 7.2 \\ 
15+14 & 1750 & 5.42 & 4.92 & 0.5 & 5.05 & 3.52 & 4.64 & 3.57 & 6.6 \\ 
15+14 & 2000 & 5.80 & 4.94 & 0.86 & 5.05 & 3.51 & 4.64 & 3.56 & 6.7 \\ 
15+14 & 2500 & 6.84 & 4.94 & 1.9 & 5.04 & 3.49 & 4.64 & 3.56 & 4.7 \\ 
\hline
17+16 & 1500 & 5.63 & 0.00 & 0.00 & 5.13 & 3.73 & 4.99 & 4.47 & 2.5 \\ 
17+16 & 1750 & 6.00 & 5.73 & 0.27 & 5.04 & 3.56 & 5.12 & 4.52 & 5.2 \\ 
17+16 & 2000 & 6.45 & 5.79 & 0.66 & 5.15 & 3.52 & 4.84 & 3.56 & 4.6 \\ 
17+16 & 2250 & 6.93 & 5.82 & 1.10 & 5.15 & 3.51 & 4.84 & 3.56 & 4.7 \\
17+16 & 2500 & 8.13 & 5.81 & 2.32 & 5.13 & 3.50 & 4.84 & 3.55 & 16.8 \\ 
\hline
20+19 & 1500 & 6.38 & 0.00 & 0.00 & 5.23 & 4.64 & 5.14 & 4.49 & 2.9 \\
20+19 & 1750 & 6.77 & 0.00 & 0.00 & 5.24 & 3.83 & 5.15 & 4.49 & 3.4 \\
20+19 & 2000 & 7.10 & 7.02 & 0.08 & 5.19 & 4.27 & 5.14 & 4.47 & 5.4 \\ 
20+19 & 2250 & 7.68 & 7.12 & 0.46 & 5.15 & 3.53 & 5.12 & 4.48 & 3.1 \\ 
20+19 & 2500 & 8.61 & 7.21 & 1.40 & 5.28 & 3.51 & 5.07 & 3.55 & 3.2 \\
\hline
\end{tabular}
\end{center}
\caption{Details of the final state of the models. These models have all been computed with a mass transfer efficiency of 0.5. The columns are: the initial masses of the stars in the system in solar masses; initial period of the system in days; the mass of the primary at explosion in solar masses; the hydrogen exhausted core mass of the primary at explosion in solar masses; the total remaining mass of the hydrogen envelope of the primary at the point of explosion, in solar masses; the logarithm of the luminosity of the primary at the point of explosion; the logarithm of the effective temperature of the primary at the point of explosion; the logarithm of the luminosity of the secondary at the point of explosion; the logarithm of the effective temperature of the secondary at the point of explosion and the N/C ratio in the ejecta material.}
\label{tab:TideMasses}
\end{table*}

Table~\ref{tab:TideMasses} gives the final state of the systems evolved. We note that the two shortest period systems of the 12+11\ms\ binary end up with both stars filling their Roche lobes at the same time. This system would form some sort of contact binary which would require additional physics (e.g. for heat transfer between the two stars) that is not included in the code. We therefore cannot comment on the outcome of these two systems. For the remaining systems, we find that at the point of explosion, the primary is barely luminous enough to fit within the observed error bars and is also rather cooler than the observations suggest. Similarly, the secondaries of these systems are also towards the lower end of the errors in the observed luminosity. This suggests that the secondaries are not massive enough and have not accreted sufficient mass during the mass transfer event. The fit could perhaps be improved by assuming the mass transfer is more than 50\% efficient. This is discussed further in section 3.2 below. As we increase the initial orbital period from 2000 days up to 2500 days, the remaining mass in the envelope increases from 0.66\ms\ to 1.90\ms\ as the binary widens, with Roche lobe overflow setting in later and hence leading to less material being stripped.

The 15+14\ms\ systems look more promising. The primaries all end up close to the observed progentior in terms of both temperature and luminosity (as can be seen in Figure~\ref{fig:p15s14}). The secondaries in the systems with initial periods below 2000 days also lie closer to the observed location of the secondary, with luminosities lying within the observed errors. However, if we have a system with an initial orbital period of 2250 days or greater, the secondary is able to evolve to a red giant before the primary explodes. This is because the wider separation means the primary has to expand more in order to fill its Roche lobe. In the time it takes for this to happen, the secondary has enough time to evolve off the main sequence. Such systems do not reproduced the observations of the system. The remaining hydrogen envelope mass of the primary varies between 0.07 and 0.65\ms, with the model that has an initial period of 2000 days having a remaining envelope mass of 0.19\ms.

The 17+16\ms\ systems show the best general agreement with the location of the progenitor system in the HR diagram. The secondaries of all but the system with the longest initial period have almost the observed luminosity. However, they are all too hot by about 0.2 dex in $\log T_\mathrm{eff}$. This suggests that the initial mass of the secondary was probably slightly higher so that it would have evolved up to (or beyond) the end of the main sequence by the time the mass transfer took place. The remaining mass in the envelope of the primary at the time of explosion varies from 0.01 to 0.44\ms\, with the model with an initial period of 2250 days having 0.20\ms\ of envelope remaining at the point of explosion.

For the highest mass pairing, 20+19\ms, we find that mass transfer is too effective at stripping the primary's envelope. Our shortest period systems end up having their surfaces completely stripped of hydrogen and so they are Wolf-Rayet stars by the time they reach the point of explosion. Only the model with an initial period of 2500 days comes close to matching the luminosity and temperature of the observed progentior. In addition, the secondaries are also too luminous and too hot. The former problem could be solved by reducing the efficiency of the mass transfer so that they accrete less mass. This should make the secondary less massive and hence less luminous.

The constraints of \citet{1994ApJ...429..300W} allow us to rule out several of the models produced above. Their modelling of the SN light curve required a model that has a final hydrogen mass of around 0.2\ms. The 12\ms\ progenitors all have final envelope masses that are too high and so can be ruled out. Of the 15+14\ms\ systems, the two shortest period systems have too little envelope remaining at the point of explosion, while the two longest period systems have too much left. The model with an initial orbit of around 2000 days is a good match to the Woosley et al. value. The 17+16\ms\ system with an initial orbit of 2250 days is also a reasonable match. The final envelope  masses of the 20\ms\ systems are all too low.

\subsection{Accretion on to the secondary}
In the above models, we assumed that only half of the material transferred from the primary was accreted by the secondary. We also noted that in the case of the 12+11\ms\ and 15+14\ms\ systems, the secondary is under-luminous compared to the observations and is therefore presumably less massive than is required. We therefore increased the value of the accretion efficiency to 1, and re-ran the models of these systems. 

\begin{figure}
\includegraphics[width=\columnwidth]{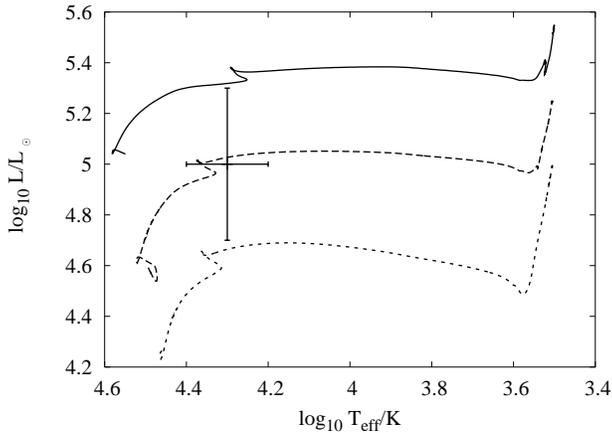}
\caption{HR diagram showing evolution tracks for a 14\ms\ (dotted line), 19\ms\ (dashed line) and 28\ms\ (solid line) star. The observed location of SN1993J's secondary is denoted by the cross with error bars. A secondary of around 19\ms\ at the time of the primary's explosion should give a good match to the observations.}
\label{fig:secondarylimit}
\end{figure}

In the case of the 12+11\ms\ systems, the enhanced accretion efficiency meant that the secondary filled its Roche lobe during the mass transfer. The code is not equipped to deal with these situations and so we terminated the simulations. Even if we could model this phase of evolution it is unlikely that we could form a secondary with a mass sufficiently large to give it a luminosity of about $10^5$\ls\ as this would require the secondary to have a mass of around 20\ms. There is not enough mass in the system to achieve this and still have a progenitor with a mass that would fit the observations. However, it would be possible to get a system that falls within the error bars of the observation, as the lower bound would require only 3\ms\ to be accepted by the secondary. This is illustrated in Figure~\ref{fig:secondarylimit}. Here, we plot evolutionary tracks for stars of 14, 19 and 28\ms. A star of less than 14\ms\ does not pass through the error bars of the observations, nor does one of 28\ms\ or greater. We note however, that these limits may change slightly when the secondary has accreted He-rich material from its companion as the difference in the opacity of this material will shift the star's position in the HR diagram (however, see section~\ref{sec:th}).  

In the case of the 15+14\ms\ runs, the two smallest initial periods resulted in the secondaries filling their Roche lobes. In the case of the wider period systems, it is possible to produce a secondary with approximately the right luminosity. However, we note that increasing the mass transfer efficiency by a factor of two has little effect on the luminosity. Despite increasing the final mass of the secondary from about 16\ms\ up to about 20\ms, we have only increased the luminosity by about 25 per cent.

The secondaries of the 17+16\ms\ systems are of about the right luminosity, so we do not vary the accretion efficiency for these models. When we look at the 20+19\ms\ systems, we note that the secondaries are too luminous. If we want to reproduce the SN1993J system starting from such a high initial mass, we need the accretion efficiency to be lower than 0.5 and perhaps the mass transfer has to be completely inefficient as a 19\ms\ star is of almost the right luminosity at the point that the primary fills its Roche lobe (see Fiugre~\ref{fig:secondarylimit}). However, the simulations show that completely inefficient mass transfer does not work: while the primary is in the core helium burning phase, the secondary crosses the Hertzsprung gap and begins to ascend its red giant branch. We would therefore need a slightly less massive secondary or a more massive primary in order for this system to work.

\begin{figure}
\includegraphics[angle=270,width=\columnwidth]{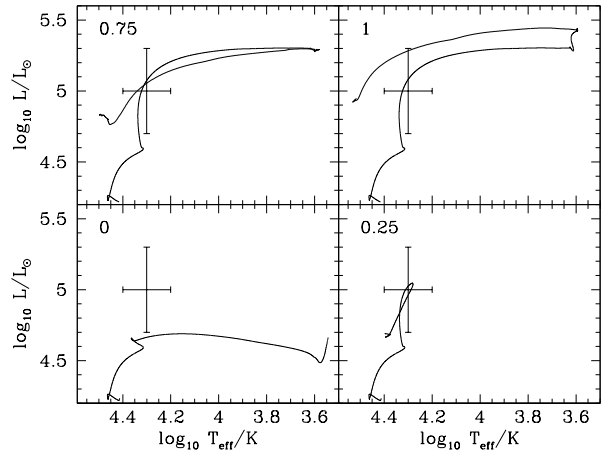}
\caption{HR diagrams showing the position of the secondary after accretion in a 15+14\ms\ system with an initial period of 2000 days. Each panel displays a track computed with a different accretion efficiency, which is marked in the top left corner of the panel. The location of SN1993J's secondary is also displayed in each panel.}
\label{fig:15qvaried}
\end{figure}

We have also examined the effect of varying the accretion efficiency for a fixed binary mass and initial separation. The evolutionary tracks for the secondary in a 15+14\ms\ system with an initial orbit of 2000 days are shown in Figure~\ref{fig:15qvaried}. If the accretion efficiency is above 0.25 then the secondary has a luminosity (and temperature) that are just consistent with the error bars of the observations. If the efficiency is 1 the secondary has about the right luminosity to match the observations, but the star is too blue. We note that if accretion is perfectly inefficient then the secondary actually becomes too evolved -- accretion is need to slightly rejuvenate the secondary.

We have a similar picture for a 17+16\ms\ system with an initial period of 2250 days. A complete spread of accretion efficiencies are consistent with the observed error bars, with an accretion efficiency of 0.5 being the best match to the observed luminosity. As with the 15+14\ms\ system, we require there to be some accretion to keep the secondary at higher temperatures and prevent it from evolving to the red.

\subsubsection{The effect of thermohaline mixing}\label{sec:th}
Because the material received by the secondary has been processed by nuclear reactions it has a higher mean molecular weight than the material in the envelope of the secondary. We therefore expect thermohaline convection to mix this material into the secondary,  altering the surface composition. This has not been taken into account in the above simulations.

As a test case, we have re-run the 17+16\ms\ model with an initial orbital period of 2000 days. This time, thermohaline mixing has been included throughout the run. The mass transfer is sufficiently rapid that thermohaline mixing is unimportant during the mass transfer phase and it is only once this phase is over that the effects of thermohaline mixing become evident. Without thermohaline mixing, the secondary has a final surface hydrogen abundance of 0.53 by mass fraction, while its helium abundance is   0.42. With the inclusion of thermohaline mixing, the final surface H abundance is 0.61 and the final surface helium abundance is 0.35. The final position of the secondary in the HR diagram is not greatly affected. With the inclusion of thermohaline mixing, the secondary is shifted to cooler temperatures by just $\Delta \log T_\mathrm{eff} = 0.02$, while the luminosity remains unchanged. We therefore do not consider thermohaline mixing any further.

\subsection{The effect of tides}

\begin{figure}
\begin{center}
\includegraphics[width=\columnwidth]{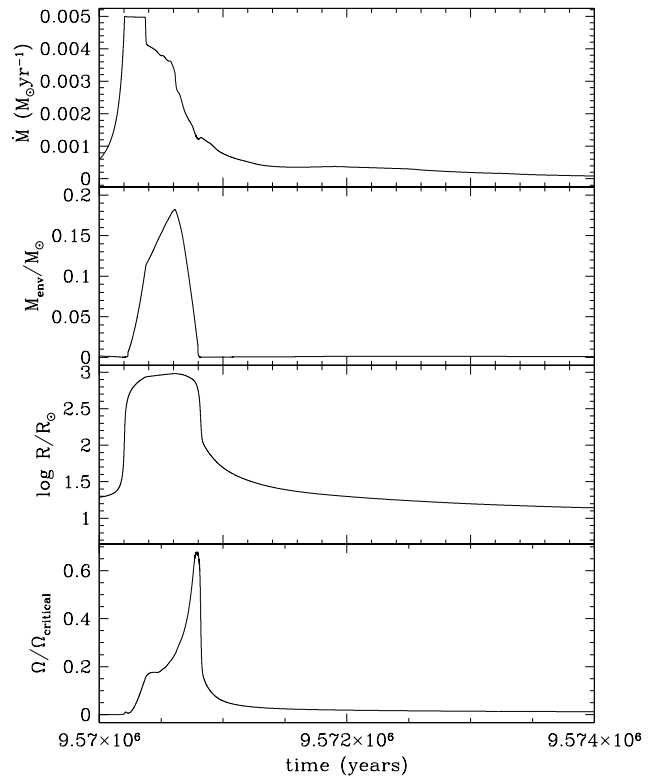}
\end{center}
\caption{Plot of the characteristics of the secondary of a system consisting of a 17\ms\ primary and 16\ms\ secondary during the phase of mass transfer. From top to bottom, the panels display: the rate of mass gain by the secondary, the mass contained in the secondary's convective envelope in solar masses, the logarithm of the stellar radius in solar radii and the rotation rate as a fraction of the critical rotation rate. The rapid accretion causes the star to swell up, temporarily becoming a giant with a convective envelope. Tidal interaction at this point increases the spin angular momentum and as the star begins to contract it spins up to an appreciable fraction of its critical rotation rate.}
\label{fig:Rotation}
\end{figure}

In the above simulations, we have included the effects of tidal interaction. As the binaries we are considering are relatively wide, we find tidal effects only become important as the stars become giants and develop deep convective envelopes. This has a particularly important effect on the secondaries of many of the systems. During the initially rapid phase of mass transfer, the secondary swells up, temporarily becoming a giant. While the mass contained in its convective envelope is small, the envelope has a large radius and so the convective turnover timescale increases sufficiently to allow tidal interaction to influence the system. The spin angular momentum of the secondary is increased and as the star contracts again it spins up, achieving an appreciable fraction of its critical rotational velocity. Figure~\ref{fig:Rotation} displays this at work in a 17+16\ms\ binary with an initial period of 1750 days.

\begin{figure}
\includegraphics[width=\columnwidth]{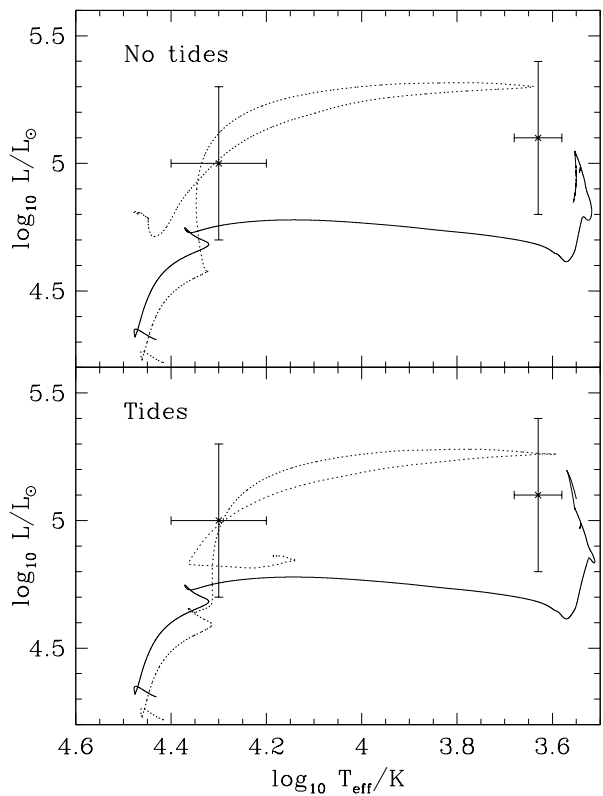}
\caption{HR diagrams showing the evolution of a 15+14\ms\ system evolved with and without tides (lower and upper panel respectively). The solid line represents the primary and the dotted line represents the secondary in each panel. The crosses with errorbars denote the observed locations of the primary (at $\log T_\mathrm{eff}/\mathrm{K}=3.63$) and secondary of SN1993J. In the upper panel, the system was started with an orbit with an initial period of 1500 days, while in the lower panel the initial period of  the orbit was 2000 days.}
\label{fig:TideComparison}
\end{figure}

We have constructed a set of models without the inclusion of tides. The details of the final states of the systems are shown in Table~\ref{tab:TideMasses}. Unsurprisingly, the models without tides have much higher final masses than the models with tides, for a given initial period. Without tides, the two stars do not get pulled together as the primary ascends the giant branch and so a smaller initial period is required to get Roche lobe overflow at a given stellar radius. Similarly, the models without tides have larger masses of hydrogen remaining at the point of explosion. The inclusion of tides is equivalent to reducing the initial period of the system by around 500 days for all the mass pairings modelled. For example, in the case of a 15+14\ms\ binary, a model without tides with an initial period of 1500 days produces a primary with similar characteristics to a model including tides with an initial period of 2000 days. These cases are shown in Figure~\ref{fig:TideComparison}. In the model without tides, the secondary has not had time for it to evolve off the main sequence before the primary fills its Roche lobe. This results in the secondary being rejuvenated by the accretion so it remains blue.

\subsection{Convective overshooting}

All the above models have been evolved using convective overshooting as mixing beyond the formal Schwarzchild convective boundary is needed to reproduce observations such as the  width of the main sequence in clusters \citep[see e.g.][and references therein]{1997MNRAS.285..696S}. The mechanism for this mixing does not have to be actual convective overshooting and could be caused by other physical mechanisms such as rotational mixing. We therefore need to examine how changing this extra mixing changes the outcome of the system. To do this, we have re-run the model grid without the use of convective overshooting.

The details of the final states of the systems are shown in Table~\ref{tab:noOS}. If we do not include overshooting, we find that the 12+11\ms\ systems do not undergo Roche lobe overflow for any of the periods we have considered. Without overshooting, the stars have smaller core masses and hence are less luminous, with smaller radii. While the final remnant masses tend to be similar to their cousins with overshooting, they have much smaller final He core masses. The non-overshooting models lose less of their envelopes, retaining larger amounts of H at the point of explosion.

\begin{table*}
\caption{Details of the final state of the models evolved without convective overshooting. These models have all been evolved with a mass transfer efficiency of 0.5. The columns are as in Table~\ref{tab:TideMasses}}
\begin{center}
\begin{tabular}{ccccccccccccc}
\hline
Initial & Initial & M$_\mathrm{final}$ & M$_\mathrm{He}$ & Envelope  & $\log$\ L$^1$/\ls & $\log$\ T$^1_\mathrm{eff}/\mathrm{K}$ & $\log$\ L$^2$/\ls & $\log$\ T$^2_\mathrm{eff}/\mathrm{K}$ & N/C \\ 
masses & period & & & mass  \\
(\ms) & (days) & (\ms) & (\ms) & (\ms) & & & & & & \\
\hline
15+14 & 1500 & 4.29 & 4.11 & 0.17 & 4.88 & 3.55 & 4.53 & 3.57 & 19.2 \\ 
15+14 & 1750 & 4.45 & 4.18 & 0.26 & 4.84 & 3.53 & 4.53 & 3.57 & 12.5 \\ 
15+14 & 2000 & 5.00 & 4.19 & 0.81 & 4.86 & 3.50 & 4.53 & 3.57 & 12.8 \\ 
15+14 & 2250 & 5.76 & 4.19 & 1.57 & 4.88 & 3.49 & 4.53 & 3.57 & 12.3 \\ 
15+14 & 2500 & 12.79 & 4.19 & 8.60 & 4.87 & 3.52 & 4.53 & 3.55 & 3.6 \\ 
\hline
17+16 & 1500 & 4.80 & 4.76 & 0.04 & 4.99 & 3.62 & 4.93 & 4.46 & 5.7 \\ 
17+16 & 1750 & 5.11 & 4.72 &  0.38 & 4.85 & 3.53 & 4.71 & 3.57 & 8.7 \\ 
17+16 & 2000 & 5.47 & 4.77 &  0.69 & 4.98 & 3.51 & 4.71 & 3.56 & 7.9 \\ 
17+16 & 2250 & 6.17 & 4.80 &  1.37 & 4.94 & 3.50 & 4.71 & 3.56 & 7.2 \\ 
17+16 & 2500 & 6.74 & 4.80 &  1.94 & 4.97 & 3.49 & 4.71 & 3.56 & 7.4 \\ 
\hline
20+19 & 1500 & 5.49 & 5.41 &  0.08 & 5.05 & 3.60 & 5.15 & 4.51 & 7.1 \\ 
20+19 & 2000 & 6.10 & 5.66 &  0.43 & 5.11 & 3.53 & 4.92 & 3.56 & 7.4 \\ 
20+19 & 2250 & 6.48 & 5.75 &  0.73 & 5.12 & 3.52 & 4.92 & 3.56 & 17.4 \\ 
20+19 & 2500 & 7.12 & 5.79 &  1.33 & 5.05 & 3.51 & 4.92 & 3.56 & 13.4 \\ 
\hline
\end{tabular}
\end{center}
\label{tab:noOS}
\end{table*}

The models without core overshooting spend less time on the main sequence and the end to the main sequence is further to the blue. This means that in order to get the secondary at the right location it {\it must} be a post-main sequence object (as opposed to being close to the end of the main sequence). This places a very serious constraint on the possible systems that can produce SN1993J -- namely both stars must have evolved beyond the main sequence in order to reproduce the observations. This is extremely difficult to achieve given that the time for crossing the Hertzsprung gap is very short. The location of the secondary in the HR diagram should be considered as evidence for the existence of extra mixing on the main sequence.

\section{Discussion}
We find we can adequately reproduce the observed parameters of SN1993J using systems with different initial masses and differing accretion efficiencies. A 15+14\ms\ system with an initial period of 2100 days will produce both a primary and a secondary in the right part of the HR diagram if the accretion efficiency is 100 per cent. This model is in very good agreement with the model presented by \citet{2004Natur.427..129M}. We obtain a similar primary mass (5.5\ms\ compared with 5.4\ms\ in the Maund et al. model), though our He-core mass is lower (4.9\ms\ versus 5.1\ms) and our secondary is also of somewhat lower mass (18.8\ms\ as opposed to 22\ms). Such a model will also have a final envelope mass of 0.55\ms\ which is not inconsistent with the observed nature of the supernova, though this is somewhat higher than the remaining envelope mass suggested by \citet{1994ApJ...429..300W}. Higher mass systems where the efficiency of mass transfer is reduced can also reproduce the observations. We find that a 17+16\ms\ system in an initial orbit of 2360 days will fit the observations provide the mass transfer is 50\% efficient. In both these cases, the secondary has evolved off the main sequence and would be observed as a blue supergiant. There is observational evidence that such systems might be quite common, as discussed by \citet{2008A&A...479..541H}.

The inclusion of tidal physics does not play such an important role in determining the final outcome of the systems. Regardless of the inclusion of tides, we still need a variation in the mass transfer efficiency in order to make the secondary of the correct luminosity. 

While comparing the luminosity and surface temperature of the models to
those dervived from observations is one way to determine the quality of
our fit, an alternative is to calculate a synthetic spectrum for each
binary system and compare these to the colours observed in the ground
based pre-explosion images. To do this we use a similar method to that described in \citet{2007MNRAS.376L..52E}. We use the BaSeL synthetic spectra library \citep{1999ASPC..192..203W} to provide spectra for each star in our binary. We chose the spectra with the surface temperature and gravity closest to that of our model star. We
then calibrate the spectra so that the bolometric luminosity is equal to
that in the stellar evolution code for a star at 10 parsecs. We work out
colours for the UBVRI pass bands using the filter functions of \citet{1990PASP..102.1181B}. We set the zero point from the flux levels given
in \citet{2000asqu.book..381D}. We have taken dust into account by using the extinction derived by \citet{2004Natur.427..129M}.

When we produce the synthetic spectral energy distributions (SEDs) for our binary systems with this method we find that the best fits are the 15 and 14\ms\ 
system with a period of 1500 days and the 17 and 16\ms\ systems
with periods of 1500 and 2000 days. The most important detail to
obtain a good match to the observed SED is the surface temparture of
the red supergiant. However this is the most uncertain variable as
small changes in the mixing length, $\alpha$, used in the prescription for convection will change the radius and surface temperature of the progenitor. Furthermore such red
supergiants may experience oscillations before core-collapse and
therefore the surface temperature may vary \citep{1997A&A...327..224H}.

The SEDs of our preferred models do generally match the observed
SED but not perfectly. For example, for the 15+14\ms\ system which had an initial period of 2100 days, we find the red supergiant surface temperature is too low with most of the flux coming out redwards of the I-band. If we increase $\log T_\mathrm{eff}$ by 0.15 dex a better fit is found. This leads us to favour models where more mass has been stripped from the primary. However, this is contrary to the SN lightcurve models of \citet{1994ApJ...429..300W}, which require a remaining hydrogen envelope of around 0.2\ms. It should also be borne in mind that the theoretical atmosphere models may not be appropriate to apply to such a highly stripped, hydrogen-poor star. 

One may question the extent to which the companion is affected by the explosion of the primary star. A substantial deposition of energy into the companion and/or stripping of material from its surface could shift its position in the HR diagram making it hotter and more luminous. The extent to which a companion is affected by a SN explosion in a binary system has been extensively studied in potential progenitor systems for Type Ia supernova \citep{2000ApJS..128..615M, 2007PASJ...59..835M}. Our progenitor systems are much more widely separated than these Type Ia progenitors and so we would expect the companion to be less affected by the explosion. There is good evidence to suggest that the companion to SN1993J did not suffer extensively during the explosion. The analysis of \citet{2004Natur.427..129M} used both pre-explosion imaging and post-explosion spectra to determine the nature of the companion \citep[see also][]{2009arXiv0903.3772M}. If the companion had been affected by the explosion they would have noted a discrepancy. Also, photometry taken in recent years \citep{2009arXiv0903.3772M} shows that the flux in the U, B and V bands (those bands in which the secondary would most likely be detected)  has returned to the same levels as observed prior to the explosion \citep{1994AJ....107..662A}. We therefore believe that the companion was not substantially affected by the primary's explosion and that its position in the HR diagram is a result of its evolution and mass-transfer history.

Additional information would be required to further constrain the model. The above assessment is based on three observable quantities, namely the location of the primary in the HR diagram at explosion, the location of the secondary in the HR diagram when the primary exploded and the final mass in the envelope of the primary at the point of explosion. The latter is based on the modelling of the supernova light curve and may be regarded as less reliable as it is an inferred quantity, rather than a directly observed one. There is also a know discrepancy between progenitor masses as determined from stellar evolution calculations and simulations of Type II SN lightcurves \citep{2008A&A...491..507U}.

\citet{2005ApJ...622..991F} measured the ratio of nitrogen to carbon in the ejecta of SN1993J. They obtained a value of $12.4\pm6$. But what does this value represent? If it represents the average N/C in the ejecta, then the 15\ms\ model falls within the error bounds, assuming we eject all the material where the binding energy exceeds $10^{44}$\,J (i.e. all the material that does not collapse to provide the energy for the supernova display). This assumes that the material is completely mixed and there is no additional nucleosynthesis in this envelope material during the ejection process. This produces 4.2\ms\ of ejected material, with an N/C ratio of 10. Applying the same criteria to the best 17\ms\ model, we get an ejected mass of 4.7\ms\ and an average N/C of 9.8. However, a 20\ms\ primary generates 6.1\ms\ of ejecta with an average N/C value of 5.4, which is just outside of the observed error bar which leads us to favour the lower mass models. The reason the more massive star has a lower N/C ratio is that, while its H-burning shell is revealed more readily by mass loss, the ejecta contains more C-rich material from below this shell.

We have confirmed that Type IIb SN can be the result of evolution in a binary system. Whether a system results in a Type IIb SN is very sensitive to the initial period. This suggests that Type IIb SN are rare. The range of periods over which models produce Type IIb SN is around 1000 days. If we assume the entire range of initial periods is 3 to 10,000 days and that the distribution is flat in $\log_{10}P$ then less than 10 percent of binaries give rise to Type IIb SNe.

Recent observations of a second Type IIb progenitor for SN
2008ax have suggested that single stars are also progenitors for some
Type IIb SNe. \citet{2008MNRAS.391L...5C} find that this SN likely had a progenitor which was a single Wolf-Rayet star with an initial mass of 27\ms. The mass range of such progenitors is small, around 1\ms. Assuming a Salpeter IMF, 1 per cent of single stars will lead to this Type of SN. Therefore binaries should be the dominant progenitors of Type IIb SNe with an upper limit on the relative fraction of these binaries being 10 per cent. This figure agrees with the recent observations of \citet{2008arXiv0809.0403S} who find that 70.6 per cent of all core-collapse SNe are Type II and of these 5.4 per cent are Type IIb. A detailed prediction of rates is beyond the scope of the small set of models presented in the paper. However, we suggest that the difficulty found in leaving just enough hydrogen on the progenitor leads to the objects being rare which is in agreement with observations.

In producing the model progenitor system we have also created a very odd blue supergiant star. It sits in the Hertzsprung gap on the HR diagram for 10 percent of the main-sequence lifetime and the surface nitrogen abundance is enhanced by up to 1 dex. Such stars could explain the unexpected number of blue
supergiants found by \citet{2007A&A...466..277H} and \citet{2008ApJ...676L..29H}. In addition to their enhanced nitrogen abundances, their extended lifetimes as supergiants could explain why so many are found in the Hertzsprung gap when few stars should exist there.

Unlike \citet{1993Natur.364..509P} we find that the companion star does evolve into a red supergiant before exploding as a SN. \citet{1993Natur.364..509P} suggested the companion of 1993J would explode as a blue supergiant and possibly be the progenitor of a SN similar to SN 1987A. We suggest this may still be possible if the companion star had begun core helium burning before it accreted any mass. In our models we find that helium burning has yet to begin and the star evolves to become a red supergiant.

\section{Conclusions}
We conclude that a binary system with a mass pairing of between 15-17 and 14-16\ms\ could have given rise to SN1993J. The lower end of this mass range is in good agreement with the model of \citet{2004Natur.427..129M}. However, there is considerable degeneracy between the initial mass pairing and the efficiency of mass transfer. For an initial mass pairing of 15+14\ms, 100 per cent efficient mass transfer is necessary to produce a secondary in the right location in the HR diagram. For a higher initial mass pairing the mass transfer efficiency can be substantially less, though we believe that it cannot be completely inefficient as some rejuvenation of the secondary is necessary to keep it to the blue. We also note that if we were to choose a model based on the luminosities and temperatures of the system at explosion (without regard to the mass of hydrogen remain as determined from lightcurve modelling), we would select models that have smaller hydrogen envelopes than the predictions of \citet{1994ApJ...429..300W}.

It proves extremely difficult to get the secondary in just the right place in the HR diagram. We do not believe that this is because the secondary was affected by the supernova explosion. The position suggests that the secondary is extremely close to (or just beyond) the end of its main sequence. We find that this can only be done in a very narrow range in initial masses and initial periods. Why should this binary supernova lie in such a restrictive region? Are we missing some important piece of the physics that would make it more likely to find companions in this location?

\section{Acknowledgements}
The authors thank the referee, Zhanwen Han, for his useful comments which have helped to improve this manuscript. They are also grateful to all those who have contributed to the development of the evolution code, thus enabling the present version to be produced. RJS is funded by the Australian Research Council's Discovery Projects scheme under grant DP0879472. He is also grateful to Churchill College for his Junior Research Fellowship during which this project began. JJE is funded by the IoA Theory rolling grant from the STFC.

\bibliography{../../../masterbibliography}

\label{lastpage}

\end{document}